\documentstyle[prd,aps,epsf,floats]{revtex}
\begin{document}
\preprint{CRSR-1115}
\draft

\title{Implementing an apparent-horizon finder in three dimensions}

\author{Thomas W.\ Baumgarte, Gregory B.\ Cook, Mark A.\ Scheel}
\address{Center for Radiophysics and Space Research, 
                Cornell University, Ithaca, New York\ \ 14853}
\author{Stuart L.\ Shapiro}
\address{Center for Astrophysics and Relativity, 326 Siena Drive,
        Ithaca, NY\ \ 14850}
\author{Saul A.\ Teukolsky\thanks{Departments
        of Physics and Astronomy, Cornell University}}
\address{Center for Radiophysics and Space Research, 
                Cornell University, Ithaca, New York\ \ 14853}

\date{\today}

\maketitle

\begin{abstract}
\widetext
Locating apparent horizons is not only important for a complete
understanding of numerically generated spacetimes, but it may also be
a crucial component of the technique for evolving black-hole spacetimes
accurately. A scheme proposed by Libson {\it et al}., based on expanding
the location of the apparent horizon in terms of symmetric trace-free
tensors, seems very promising for use with three-dimensional numerical
data sets.  In this paper, we generalize this scheme and perform a number
of code tests to fully calibrate its behavior in black-hole spacetimes
similar to those we expect to encounter in solving the binary black-hole
coalescence problem.  An important aspect of the generalization is that
we can compute the symmetric trace-free tensor expansion to any order.
This enables us to determine how far we must carry the expansion to
achieve results of a desired accuracy. To accomplish this generalization,
we describe a new and very convenient set of recurrence relations which
apply to symmetric trace-free tensors.
\end{abstract}

\widetext
\section{Introduction}
\label{sec:intro}
A major goal of numerical relativity is to simulate the coalescence
of an orbiting pair of black holes.  In studying such systems,
we will be interested in determining many quantities:  the
energy and momentum radiated, the associated waveforms, the
total angular momentum of the system, etc.  In addition to these
common physical quantities, we will also want to understand the
causal structure of the spacetime.  Not only will this give us
a more complete picture of the dynamics, but it also seems that
tracking the causal structure may prove to be a crucial step in
successfully evolving black-hole spacetimes\cite{seidel_suen92}.
Knowing which events are inside black holes may allow them to be
excised from the computational domain, thereby avoiding numerical
difficulties that have plagued black hole evolutions.  Ideally, we
would like to be able to track all of the event horizons in a
given spacetime.  However, this is not possible {\em during}
the evolution: event horizons can only be reconstructed after the 
evolution is complete.  Instead, {\em apparent horizons} can be
located on each individual space-like hypersurface during an
evolution.  Since apparent horizons must lie inside event horizons
and asymptote towards them as the system settles down, they provide
much of the desired causal information.  They can also be used to
define regions that can be excised from a computation.

Various methods exist for locating apparent horizons.  In practice,
one searches for marginally outer-trapped surfaces
(MOTS)\cite{hawkingellis}.  The apparent horizon is the outermost 
such surface.  For axisymmetric problems, shooting
methods\cite{cadez74,bishop82,bishop84,st92} have been
the most widely used for locating apparent horizons,
although decomposition into orthogonal polynomials\cite{eppley77},
the solution of elliptic boundary-value problems\cite{cook90,cookabrahams92},
and the use of curvature flows\cite{tod91} have been used.  Unfortunately,
shooting methods do not generalize to three-dimensional spatial slices.
The first general apparent-horizon finders were based on a
spherical-harmonic decomposition of the
MOTS\cite{nakamura84,nakamura85,bishop91}.  In this approach, each
coefficient in the spherical-harmonic expansion is determined,
iteratively, by performing an integral over a complicated function
that characterizes the MOTS.  The MOTS equation can also be posed
as an elliptic equation for a function that parametrically specifies
the location of the MOTS\cite{huq96}.  Curvature flow methods are
also certainly applicable in the general case of a three-dimensional
spatial hypersurface.

Recently, a variant of the spherical-harmonic decomposition method
has been proposed by Libson {\it et al}.\cite{libson95}.  This
approach is conceptually appealing, having two particularly nice features.
First, the coefficients in the spherical-harmonic expansion are
determined by a minimization procedure, eliminating the need to perform
surface integrals.  Second, Libson {\it et al}. have proposed the
use of symmetric trace-free (STF) tensors for parametrically representing
the MOTS.  This latter feature is particularly appealing when Cartesian
coordinates are used on the three-dimensional hypersurface.

In this paper, we will review the method proposed by Libson {\it et al}.
and describe how we generalize the method by extending the expansion in
STF tensors to arbitrary order.  Because one is always using a truncated
expansion, it is important to understand clearly the behavior of the
apparent-horizon finder when the maximum order of the expansion is varied.
As we will see, the number of points where the MOTS is determined will
also affect the behavior of the apparent-horizon finder.  We have examined
both of these effects in detail.

The paper is organized as follows: In \S~\ref{sec:methods} we outline the
method and basic equations, and in \S~\ref{sec:numerics} we explain our
numerical implementation. In \S~\ref{sec:tests} we carefully discuss
results from various test calculations, and in \S~\ref{sec:summary} we
briefly summarize the most important results. All technical details are
provided in the appendices. Appendix~\ref{app:expressions} contains a
number of useful equations relating to STF tensors.
Appendix~\ref{app:storage} describes the storage for arbitrary rank
tensors.  In Appendix~\ref{app:init_recur} we derive recurrence relations
for STF tensors.  Finally, in Appendix~\ref{app:area_elem} we derive an
expression for the area element on the apparent horizon.

\section{Method and basic equations}
\label{sec:methods} 
A MOTS is a closed two-surface embedded in a three-dimensional spatial
hypersurface and, therefore,  can be defined as a level surface of
some scalar function $\tau$.  If we use Cartesian coordinates on the
hypersurface then we can parametrically define the level surface as
\begin{equation}
\tau(x,y,z) = \sqrt{\delta_{ij}(x^i - C^i)(x^j - C^j)} - f(\theta,\phi) = 0.
\end{equation}
Here, $x^i$ are Cartesian coordinates, $C^i$ is a location {\em inside}
the $\tau = 0$ surface, and $\theta$ and $\phi$ are polar coordinates
centered on $C^i$. The function $f(\theta,\phi)$ then measures the
{\em coordinate} distance between $C^i$ and the $\tau=0$ surface in the
direction $(\theta,\phi)$. The outward pointing unit normal on the
$\tau=0$ surface is
\begin{equation}
S_i = \lambda \partial_i \tau,
\end{equation}
where $\lambda$ is the normalization factor
\begin{equation}
\lambda \equiv \left[ g^{ij} (\partial_i \tau)(\partial_j \tau )
						\right]^{-\frac{1}{2}}
\end{equation}
and $g_{ij}$ is the metric on the spatial hypersurface.
The expansion $\Theta$ of an outgoing null-bundle can now be written
\begin{equation}
\Theta = D_i S^i + K_{ij} S^i S^j - K^i_i,
\end{equation}
where $K_{ij}$ is the extrinsic curvature and $D_i$ is the covariant 
derivative operator associated with $g_{ij}$. Note that the term $D_i S^i$
involves both first and second derivatives of $\tau$ and hence of 
$f(\theta,\phi)$. Writing these out explicitly yields
\begin{equation} \label{theta}
\Theta = (g^{ij} - S^i S^j) \left( \frac{\lambda}{f} (\delta_{ij}
        - n_i n_j) - \lambda \partial_i \partial_j f
        - S_k \Gamma^k_{ij} - K_{ij} \right).
\end{equation}
Here $\delta_{ij}$ is the Kronecker delta, $n^i$ is the unit vector
in the $(\theta,\phi)$ direction, and $\Gamma^k_{ij}$ are the 
connection coefficients associated with $g_{ij}$.  One must be
careful in deriving this equation since there are effectively two
metrics being used: the spatial metric $g_{ij}$ and the Kronecker
delta $\delta_{ij}$ used in computing coordinate distances.  In
particular, we note that
\begin{eqnarray}
n^i &=& \frac{x^i - C^i}{\sqrt{\delta_{jk}(x^j - C^j)(x^k - C^k)}} \\
n_i &\equiv& \delta_{ij}n^j \ \ (\delta_{ij}n^in^j = 1) \\
S_i &=& \lambda(n_i - \partial_if) \\
S^i &\equiv& g^{ij}S_j \ \ (g_{ij}S^iS^j = 1)
\end{eqnarray}
 
Our goal is now to find a function $f(\theta,\phi)$ such that the 
$\tau = 0$ surface is a MOTS, i.e. that the 
expansion~(\ref{theta}) vanishes on that surface. In practice,
instead of making the expansion vanish, we can evaluate it at a number
$N_\Theta$ of points on the surface $\tau = 0$ and look for a function
$f$ such that
\begin{equation} \label{AHsum}
{\cal S}(N_\Theta) \equiv \sum_{\alpha = 1}^{N_\Theta}W_\alpha \Theta_\alpha^2
\end{equation}
vanishes.  If (\ref{AHsum}) vanishes for arbitrary weights $W_\alpha$,
then in the limit $N_\Theta\rightarrow\infty$ (so as to completely
cover the surface) we are guaranteed of having located a MOTS.
 
Our strategy will be to expand $f(\theta,\phi)$ in terms of
multipole moments and to search for a minimum in ${\cal S}$. The
sum~(\ref{AHsum}) then depends on the corresponding expansion
coefficients, which can be varied until the sum assumes a minimum.
If this minimum comes arbitrarily close to zero an apparent horizon has
been found. Thus, the problem has been reduced  to a multidimensional
minimization, for which standard methods can be used.
 
An obvious choice of basis functions for the expansion of $f(\theta,\phi)$
are the spherical harmonics,
\begin{equation} \label{sph}
f(\theta,\phi) = \sum_{\ell=0}^{L} \sum_{m=-\ell}^{\ell} F^{\ell m}
	Y^{\ell m}(\theta,\phi),
\end{equation}
where the expansion is truncated at order $L$. However, since we have
to take up to second derivatives with respect to Cartesian coordinates,
an expansion in terms of STF tensors 
\begin{equation} \label{stf}
f(\theta,\phi) = \sum_{\ell=0}^L {\cal F}_{K_\ell} N_{K_\ell}
\end{equation}
turns out to be a better choice. In the following we adopt the notation
of~\cite{thorne80}, where additional details of this formalism can be found. 
In particular, repeated indices will always be summed over. The 
subscript $K_\ell$ denotes a multi-index of length $\ell$, and $N_{K_\ell}$
is the vector product of $\ell$ unit vectors $n_i$:
\begin{equation} \label{nkl}
N_{K_\ell}= n_{k_1} n_{k_2} \cdots n_{k_\ell}.
\end{equation}
In~(\ref{stf}), these are contracted with the STF tensors ${\cal F}_{K_\ell}$
(of rank $\ell$). These are the {\em location-independent} expansion
coefficients equivalent to the $F^{\ell m}$ in~(\ref{sph}).  Note that
a STF tensor of rank $\ell$ has $2\ell+1$ independent components, just
like the spherical harmonics.  The relationship between~(\ref{sph})
and~(\ref{stf}) can be seen even more clearly by choosing the
${\cal Y}^{\ell m}_{K_\ell}$ basis for the STF tensors as defined
in Ref.~\cite{thorne80} (see Appendix~\ref{app:expressions}). In terms
of these, ${\cal F}_{K_\ell}$ can be written
\begin{equation} \label{sum}
{\cal F}_{K_\ell} = \sum_{m=-\ell}^{\ell} F^{\ell m}
				{\cal Y}^{\ell m}_{K_\ell},
\end{equation}
where the $F^{\ell m}$ are the same as in~(\ref{sph}). The 
${\cal Y}^{\ell m}_{K_\ell}$ also provide a relation between the
spherical harmonics and the $N_{K_\ell}$
\begin{equation} \label{ylm}
Y^{\ell m}(\theta,\phi) = {\cal Y}^{\ell m}_{K_\ell} N_{K_\ell}(\theta,\phi).
\end{equation}
Inserting this into~(\ref{sph}) and using~(\ref{sum}) 
immediately yields~(\ref{stf}).
 
Note that because the coefficients ${\cal F}_{K_\ell}$ are location
independent, derivatives of $f$ can  be calculated from derivatives
of $N_{K_\ell}$
\begin{equation}
\partial_i f(\theta,\phi) = \sum_{\ell=0}^L {\cal F}_{K_\ell}
						\partial_i N_{K_\ell}
\end{equation}
and similarly for second derivatives.

\section{Numerical Implementation}
\label{sec:numerics} 
Our code is designed in such a way that it can find a MOTS to arbitrary
order $L$ in the multipole expansion.  On input we therefore have to
specify the order $L$.  Also, we have to specify the number of points
$N_\Theta$ on the surface at which the expansion~(\ref{theta}) (and,
of course the sum for ${\cal S}$) are evaluated. These points must be
distributed somehow over the surface.  Currently, they are distributed
equally in $\phi$ and $\cos\theta$ on the unit sphere, but different
choices could easily be made.  Results for different values of $L$ and
$N_\Theta$ will be presented in the next section.
 
Next the tensors $N_{K_\ell}$, their first and second derivatives, as
well as the basis STF tensors ${\cal Y}^{\ell m}_{K_\ell}$ have to be
initialized. The latter are independent of location, so that we need to
calculate them only once.  The $N_{K_\ell}$ do, however, depend on the
direction $(\theta,\phi)$, and therefore have to be calculated once for
every point on the surface.  In the code, we define arrays of length
$N_\Theta$ to store $N_{K_\ell}$ and its derivatives.
 
Since all of the STF tensors are completely symmetric, we can store
the independent components in a very elegant and efficient way. This
is explained in detail in Appendix~\ref{app:storage}. In
Appendix~\ref{app:init_recur} we present recurrence relations 
that allow for a very efficient initialization of these objects.
 
The MOTS search is started with a set of trial expansion coefficients
$F^{\ell m}$. These are then contracted with the basis STF tensors
${\cal Y}^{\ell m}_{K_\ell}$, which yields the ${\cal F}_{K_\ell}$.
Since these quantities are independent of location, this needs to be
done only once per iteration step.
 
The ${\cal F}_{K_\ell}$ are then contracted with $N_{K_\ell}$
and its derivatives to find $f$, $\partial_i f$ and 
$\partial_i\partial_jf$ for each direction $(\theta,\phi)$.  Once $f$ 
is known we can construct the coordinate location
\begin{equation}
x^i = f n^i + C^i
\end{equation}
of the trial surface ($\tau = 0$) at each of the $N_\Theta$ points on
the surface. For each point, we read in $g_{ij}$, $K_{ij}$ and
$\Gamma^k_{ij}$.  These can either be numerically evolved quantities or,
for the test purposes in this paper, analytical values.  Eq.~(\ref{theta})
now yields the expansion $\Theta$ for this location on the trial surface.
Repeating these steps for every point $N_\Theta$ we can finally construct
the sum ${\cal S}$.  Currently, we choose the weights $W_\alpha$ in
(\ref{AHsum}) based on the proper area element (\ref{area_elem}) defined
in Appendix~\ref{app:area_elem}.  Thus, equation~(\ref{AHsum}) is an
approximation to the mean square of the expansion:
\begin{equation}
\label{AHint}
{\cal S} = \oint\Theta^2d^2\sigma
\end{equation}
 
Any multidimensional minimization routine can now be used to vary the
$F^{\ell m}$ until ${\cal S}$ has assumed a minimum. So far we have
found best results with Powell's method\cite{numrec_c}, although it is
likely that a method that uses derivatives with respect to the expansion
coefficients $F^{\ell m}$ will be significantly faster, especially when
the initial guess is already close to the final answer.  We hope to
explore this in the future.
 
Once a minimum has been found we shift the center of the black hole
$C^i$ according to the dipole moment (i.e.\ we choose $C^i$ so that the
$\ell=1$ moments vanish).  We then repeat the MOTS search until the
$\ell=1$  moments stay below a predetermined maximum value.  This 
procedure enables us to locate apparent horizons even when the initial
guess is very poor (see next section) and should allow us to follow black
holes that move through a numerical grid.

\section{Tests}
\label{sec:tests}
\subsection{Schwarzschild}
\label{sec:schwtest}
An obvious test for the apparent-horizon finder is the Schwarzschild
spacetime. Since the MOTS is spherically symmetric, it can be described
with the monopole term alone.
 
This test strongly demonstrated how well the shifting of the center $C^i$
according to the $\ell=1$ moments works. The code was able to locate the
MOTS accurately even when the initial guess was completely disjoint from the
true horizon. The code worked equally well when we located the black hole
away from the origin of the coordinate system. In all cases the sum
${\cal S}$, as well as all expansion coefficients $F^{\ell m}$ with
$\ell > 0$, vanished to whatever tolerance we specified.
 
\subsection{Two black holes}
\label{sec:2_bh}
A spacetime containing two black holes has multiple MOTS, some of which
can be highly distorted. Such a spacetime provides a much stronger test
for the apparent-horizon finder.
 
A metric for two time-symmetric black holes can be written in the
conformally-flat form
\begin{equation}
ds^2 = \psi^4(dx^2 + dy^2 + dz^2),
\end{equation}
where the conformal factor $\psi$ is given by
\begin{equation} \label{confact}
\psi = 1 + \frac{M}{2r_1} + \frac{M}{2r_2} 
\end{equation}
and $r_1$ and $r_2$ are
\begin{eqnarray}
r_1 & = & \left( x^2 + y^2 + (z + z_0)^2 \right)^{1/2} \nonumber\\
r_2 & = & \left( x^2 + y^2 + (z - z_0)^2 \right)^{1/2}. 
\end{eqnarray}
Here $M$ is the mass of the individual black holes, and $z_0$ is their
coordinate distance from the origin of the coordinate system. 
 
Note that the singularities in~(\ref{confact}) can be removed by adding
matter sources (see Ref.~\cite{st92}). Since this is advantageous in a
numerical application and does not change the external metric we have 
implemented this form of the equations.
 
The causal structure of this spacetime has been investigated in detail
by Bishop\cite{bishop82}. The MOTS can be found by using a shooting method
to solve a set of coupled differential equations to high accuracy.  This
provides us with a solution that we can check the STF based apparent-horizon
finder against. In the following we will refer to these solutions as the
``true horizons''.
 
In general there will be MOTS around the individual holes. If the holes
are close enough, a pair of encompassing MOTS will also appear (see
Ref.~\cite{bishop82} for a careful discussion).  According to
\v{C}ade\v{z}~\cite{cadez74} these encompassing MOTS first appear at
as separation of $z_0 = 0.765$.  For separations close to the critical
separation (i.e.\ $z_0 \lesssim 0.765$) they will be strongly distorted.
 
Note that this situation is quite similar to what we expect in
a binary black-hole evolution.  Having that application in mind, it 
provides a strong test for our apparent-horizon finder and it can help
us to decide to which order $L$ we need to expand and at how many points
$N_\Theta$ we need to evaluate the expansion $\Theta$ in order to
accurately locate the an encompassing MOTS.
 
\begin{figure}
\epsfxsize=7in\epsffile{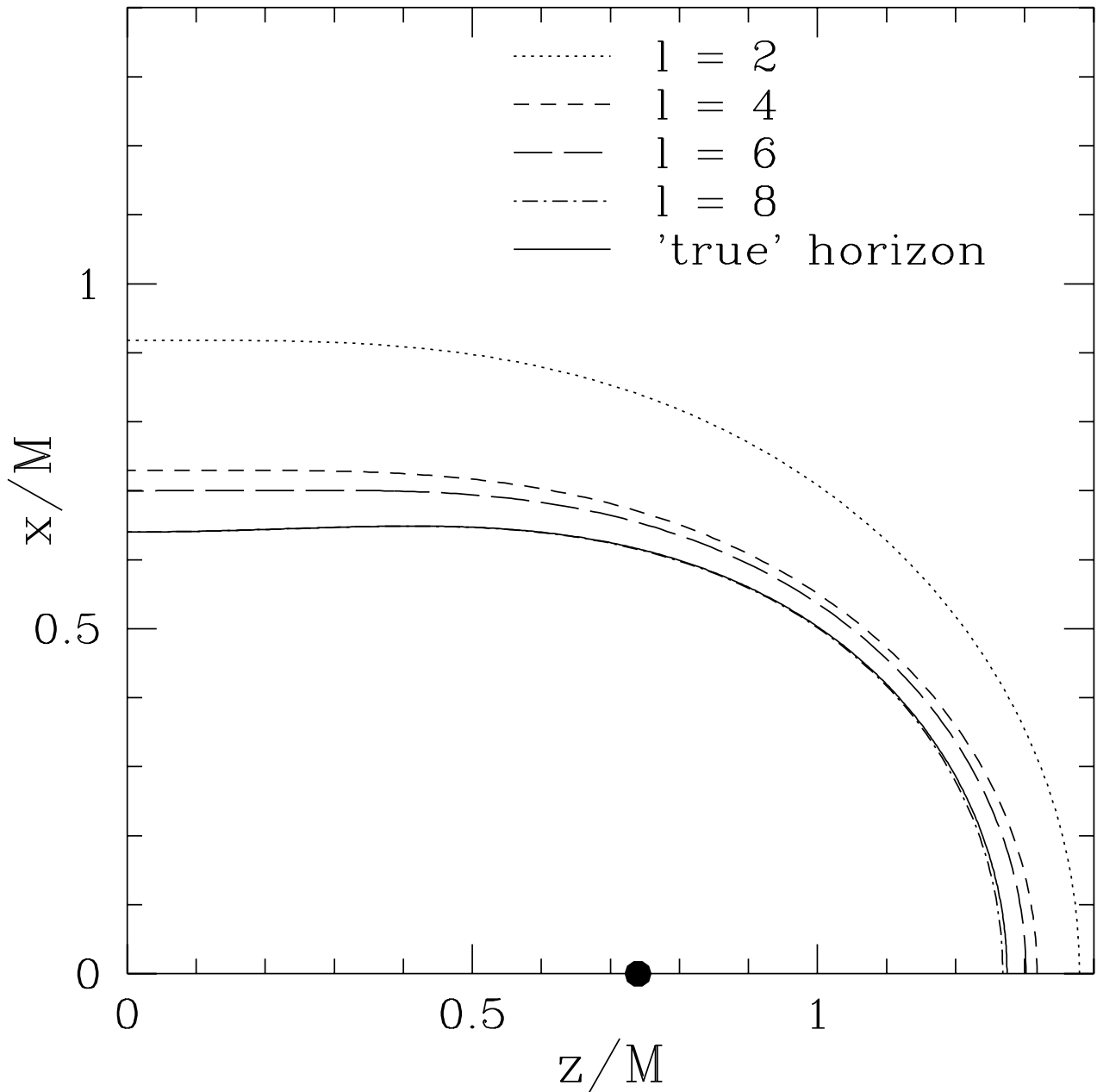}
\caption{The estimated location of the MOTS found with expansions
taken to order $L = 2$, $4$, $6$ and $8$ for $z_0 = 0.74$ (using
$N_\Theta = 25\times11$ points). The ``true location'' was found
independently by solving a set of coupled differential equations
(see Ref.~\protect\cite{st92}). The dot marks the coordinate center
 of one of the two black holes.}
\label{fig:location}
\end{figure}

In Fig.~\ref{fig:location} we plot the estimated location of the MOTS
based on expansions to order $L = 2$, $4$, $6$ and $8$ for the case
$z_0 = 0.74$.  (By symmetry, only even $L$ can contribute since
$C^i\rightarrow0$). The ``true horizon'' is fairly distorted, causing
the lower order expansions to perform very poorly. 
 
Note that all the lower order expansions find a location for the MOTS
outside of the true horizon.  This behavior could have severe consequences
in numerical evolution codes that use ``apparent horizon boundary
conditions''\cite{seidel_suen92} and that ignore the causally disconnected
region inside a MOTS.  If we were to use one of the lower-order MOTS
solutions for this purpose we could be ignoring a region that is {\em
not} causally disconnected.
  
This test clearly demonstrates that high order expansion is absolutely
necessary for the detection of highly distorted horizons.  On the other
hand it also demonstrates that high order expansion is very expensive:
increasing the order of the expansion by 2 increases the CPU time by
roughly a factor of 3, ranging from several seconds for $L = 2$ to
several minutes for $L = 8$ (on a serial computer). Details will
depend on the particular numerical implementation as well as
parameters associated with a given minimization routine.  However,
since we are searching for minima in an $(L+1)^2$-dimensional space
(see eq.~(\ref{flm})) the required CPU time will always be a steep
function of $L$.
  
\begin{figure}
\begin{picture}(504,504)
%%\put(-36,0){\framebox(504,504){}}
%%\put(-36,252){\framebox(252,252){}}
\put(0,268){\epsfxsize=3.5in\epsffile{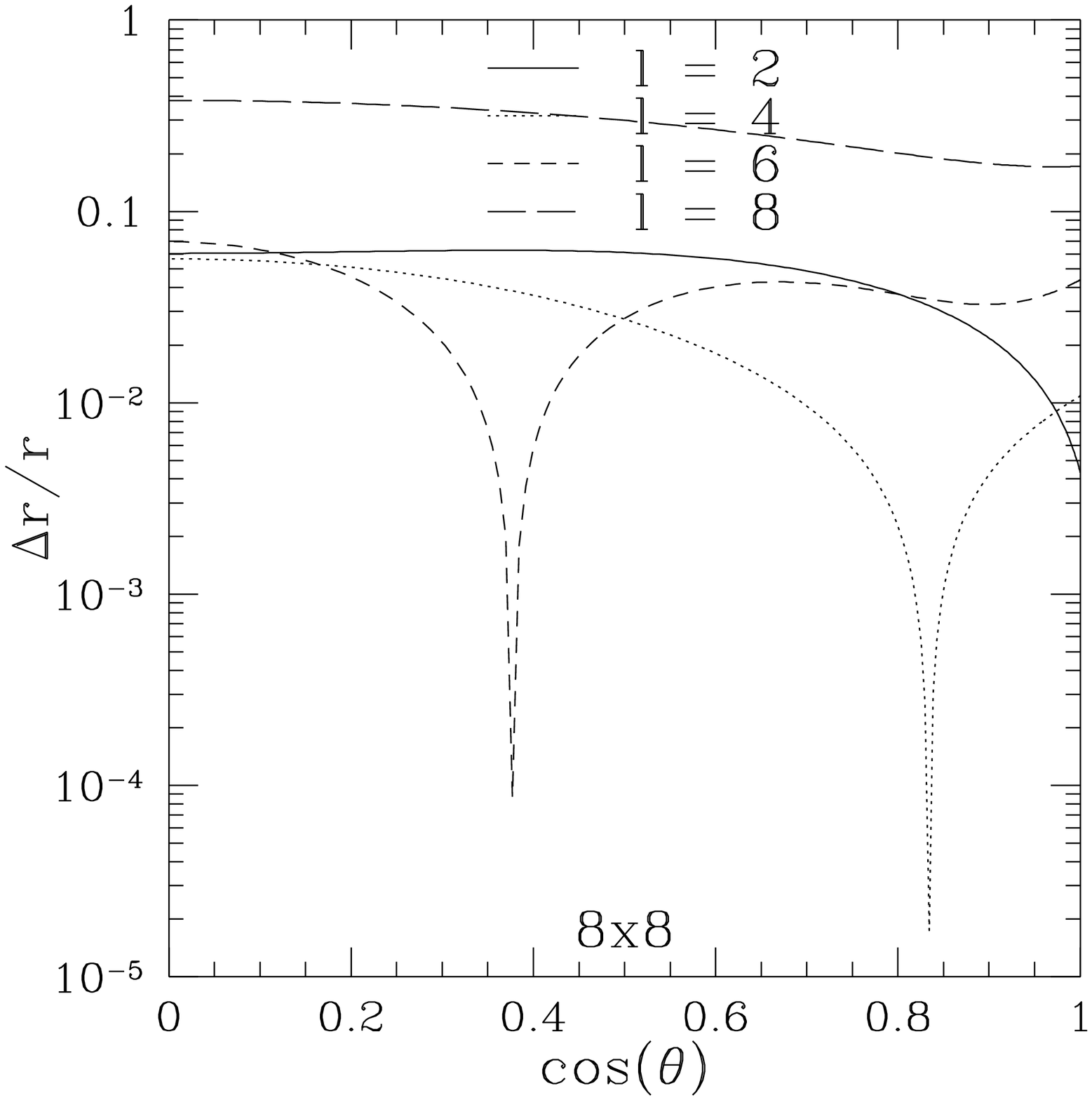}}
%%\put(216,252){\framebox(252,252){}}
\put(252,268){\epsfxsize=3.5in\epsffile{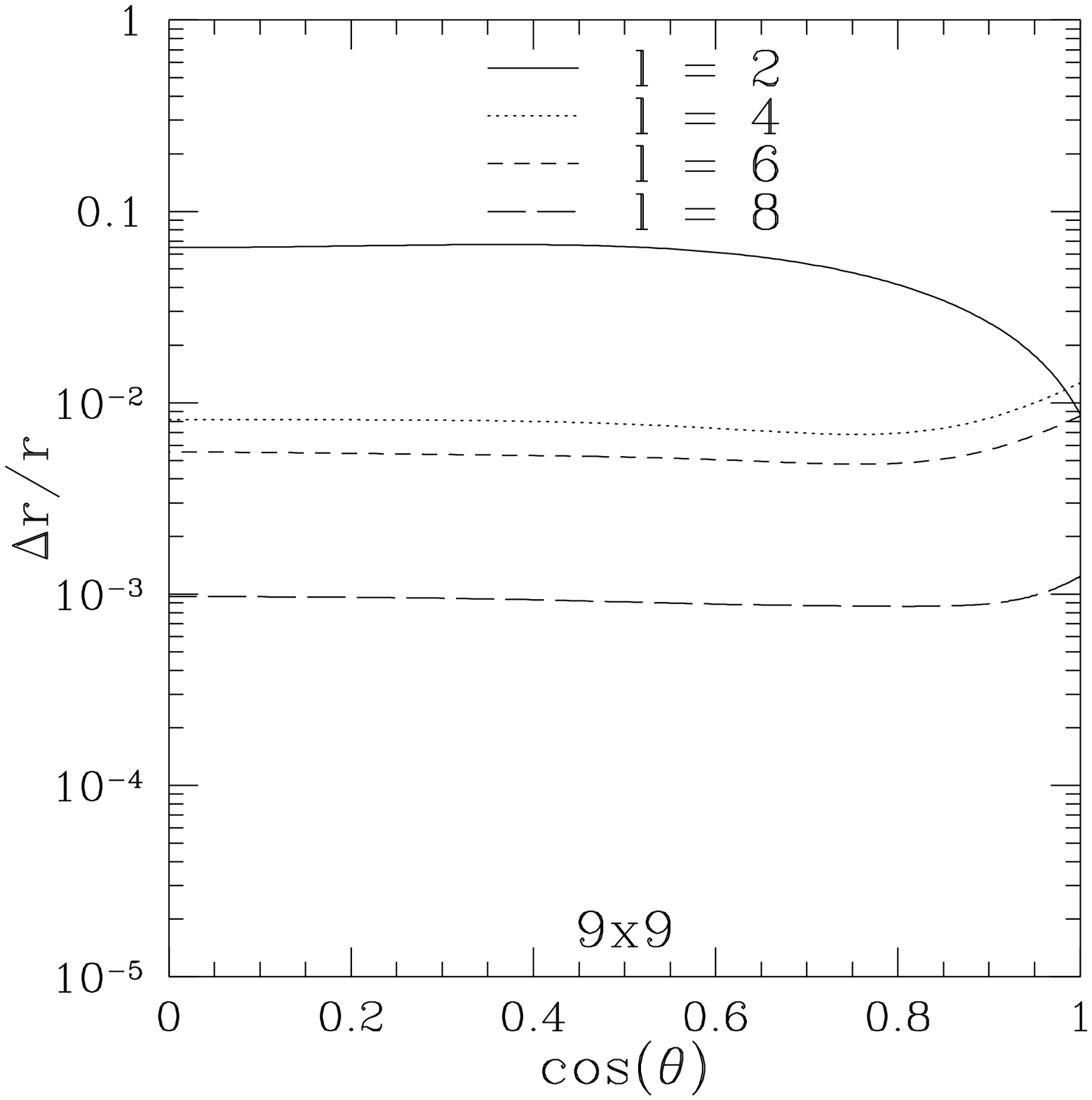}}
%%\put(-36,0){\framebox(252,252){}}
\put(0,16){\epsfxsize=3.5in\epsffile{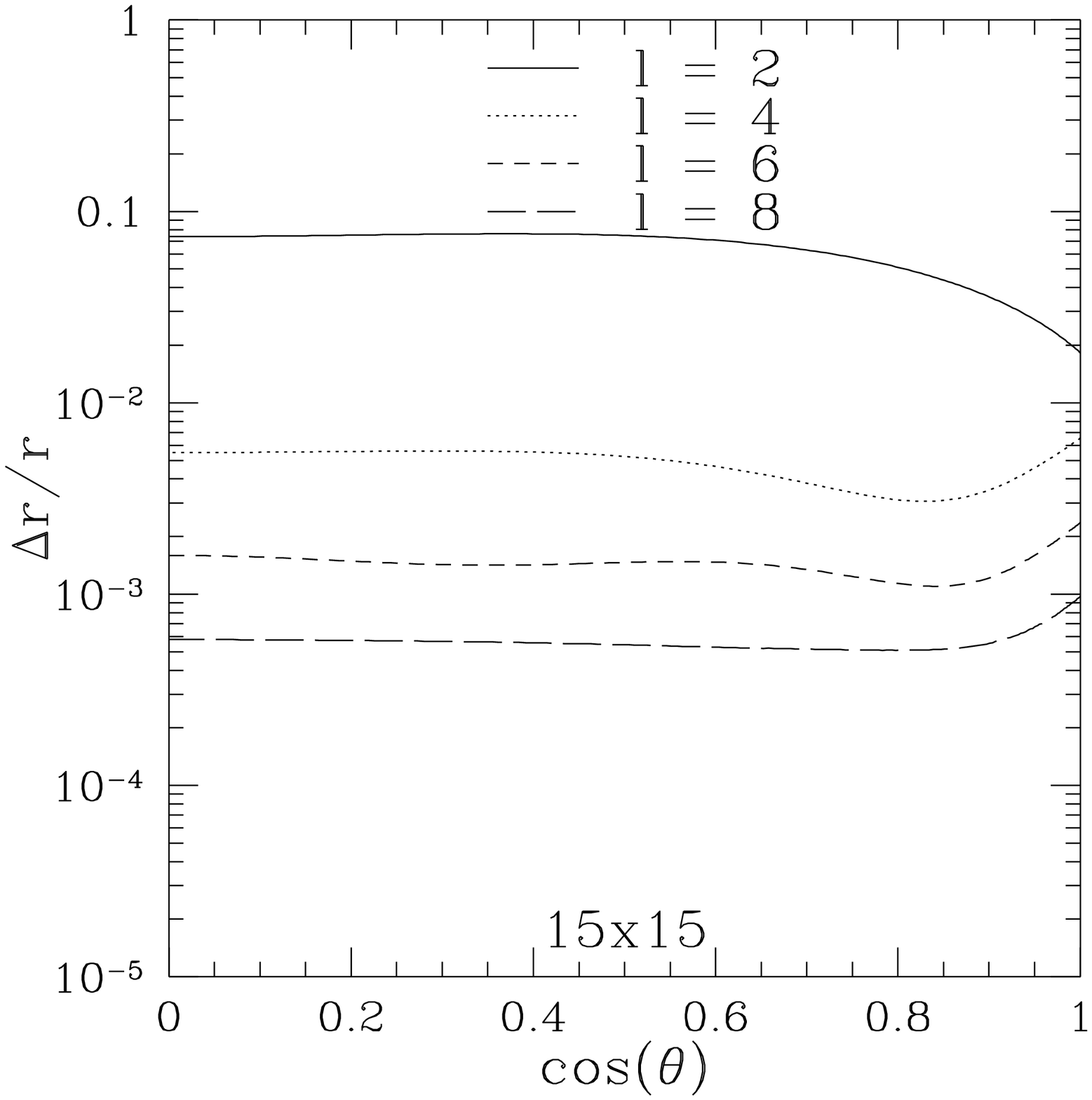}}
%%\put(216,0){\framebox(252,252){}}
\put(252,16){\epsfxsize=3.5in\epsffile{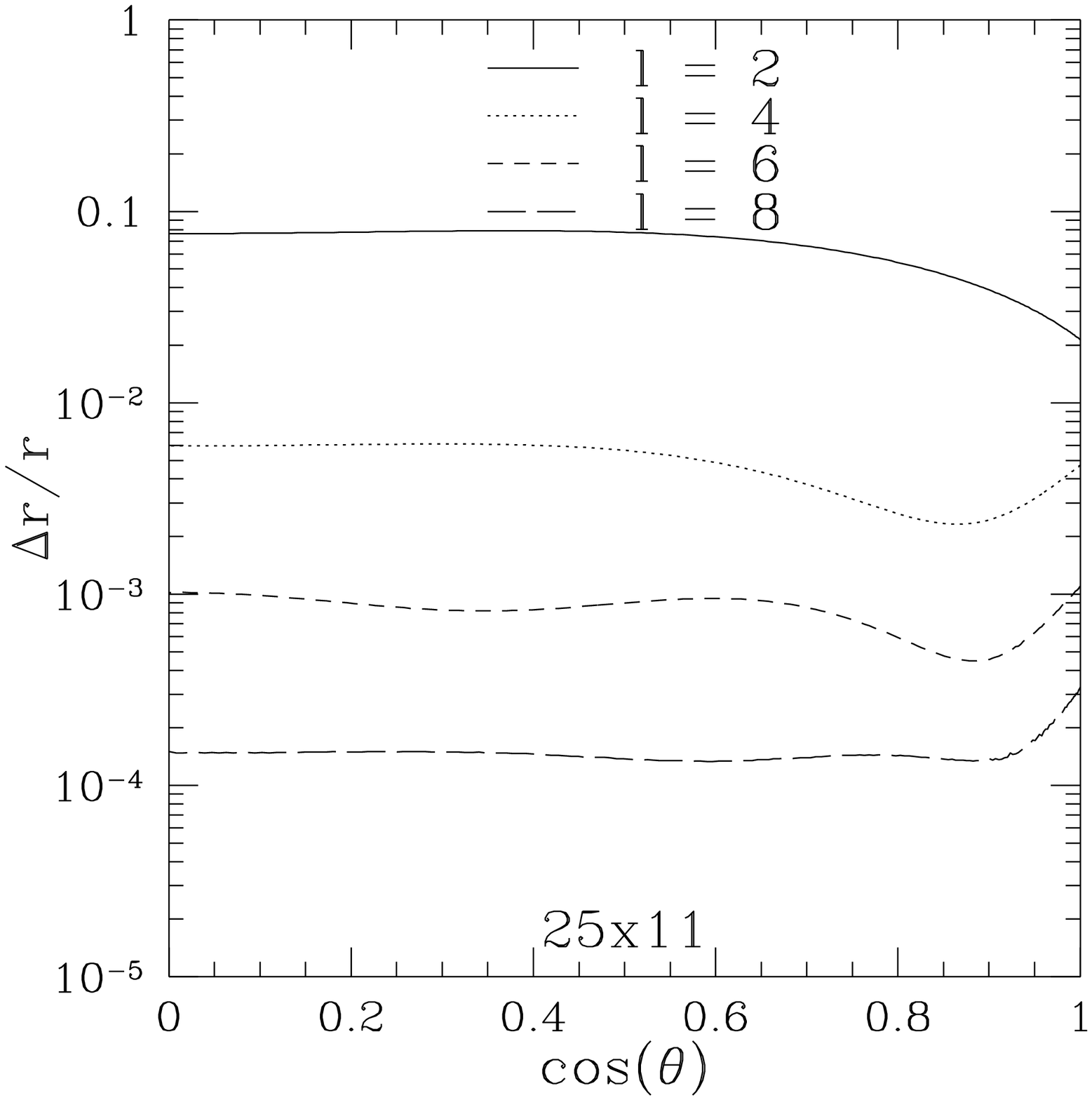}}
\end{picture}
\caption{Relative errors $\Delta r / r$ as a function of $\cos \theta$
for expansion to order  $L = 2$, $4$, $6$ and $8$ and for different 
values of $N_\Theta = n_{\theta} \times n_{\phi}$.  ($z_0 = 0.6$).}
\label{fig:errors}
\end{figure}

Another important factor for both the accuracy and the CPU time is the
number of points $N_\Theta$ at which the expansion $\Theta$ is evaluated. 
In Fig.~\ref{fig:errors} we show results for $z_0 = 0.6$ using different
numbers of points $N_\Theta = n_{\theta} \times n_{\phi}$, where
$n_{\theta}$ is the number of points in the $\theta$-direction and
$n_{\phi}$ in the $\phi$-direction.
 
Since the $F^{\ell m}$ up to order $L$ have $(L+1)^2$ independent 
components, we will need at least $(L+1)^2$ points. From
Fig.~\ref{fig:errors} it is obvious that $8\times8$ points are not
enough for an expansion to order $L=8$: the result is worse than for
a lower order expansion.  However, it can also be seen that increasing
the number of points beyond this minimum can drastically increase the
accuracy.
  
\begin{figure}
\epsfxsize=7in\epsffile{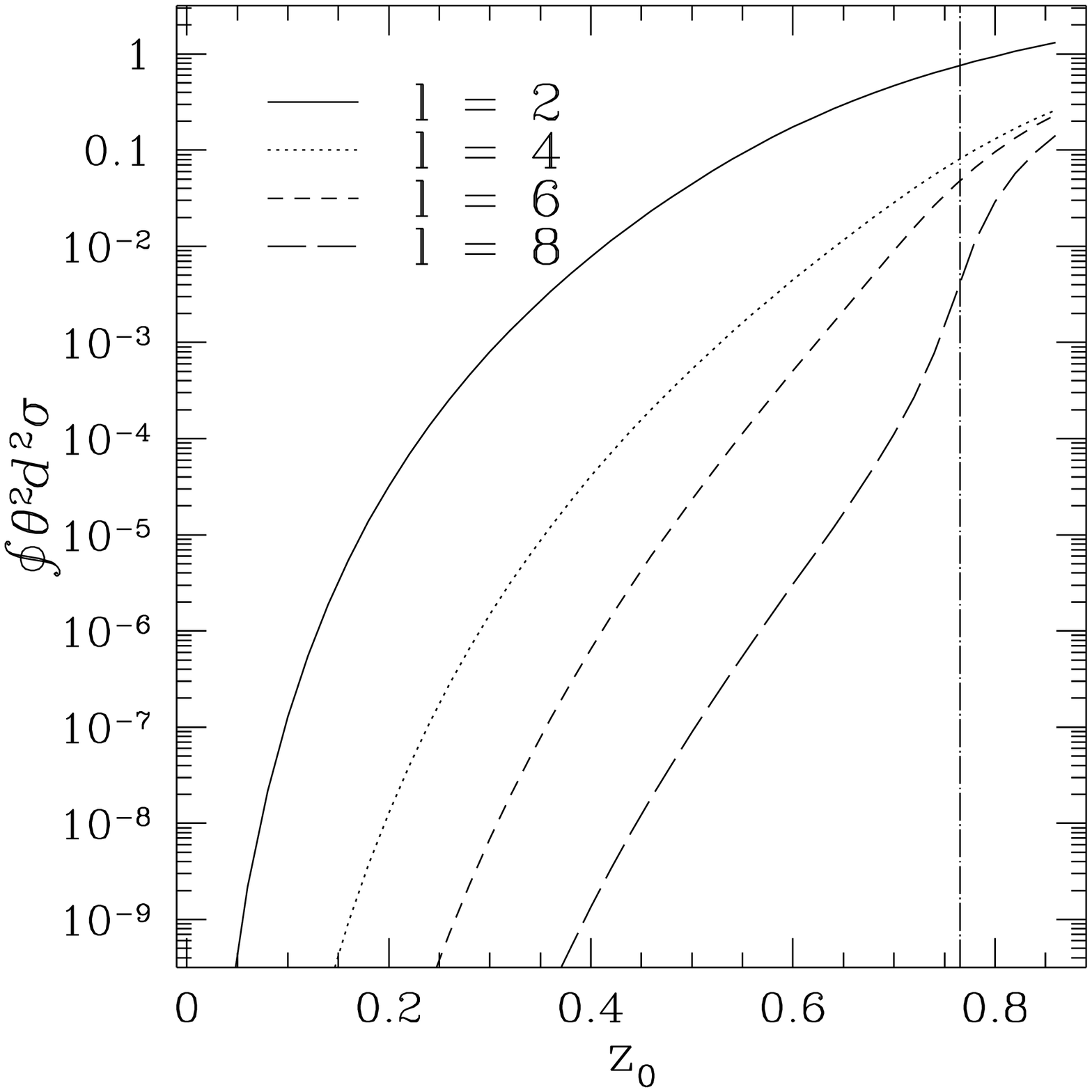}
\caption{Integrals over the expansion $\Theta$ (eq.~(\protect\ref{AHint}))
as a function of $z_0$ for expansions to order $L = 2$, $4$, $6$ and
$8$ (using $N_\Theta = 25 \times 11$ points). For $z_0 < 0.765$, i.e.
left of the vertical line, the two black holes have an encompassing MOTS.}
\label{fig:expansion}
\end{figure}

As a next test we plot in Fig.~\ref{fig:expansion} the integral
(\ref{AHint}) as a function of $z_0$ for different expansion orders.
For all these calculations we start with an initial guess close to where
we expect a common MOTS. For $z_0 < 0.765$, when the two black holes
have a common MOTS, we would therefore expect this sum to vanish, if we
could resolve the MOTS arbitrarily well. For values of $z_0$ larger than
$0.765$ the minimization routine will find a nonzero minimum -- however
we expect these minima to go to zero as we approach $z_0 = 0.765$.
 
Since we use an expansion to finite order we cannot resolve the MOTS
arbitrarily well. This means that, while we will find a minimum, it
will typically be different from zero even for $z_0 < 0.765$. The value
will, again, depend on a number of parameters, but primarily on the order
of the expansion $L$. This can be seen very clearly in
Fig.~\ref{fig:expansion}. In particular for the lower order expansions
the expected drop at $z_0 = 0.765$ cannot be detected at all -- a
significant decrease can only be seen for $L = 8$. This demonstrates
again that an early detection of a common MOTS will only be possible
with a high order expansion.
 
Also, this suggests that it is hardly possible to decide on the basis 
of the value of the sum ${\cal S}$ if a MOTS has been found. As a better
test, we suggest checking whether $\Theta$ is negative everywhere on a
surface just inside the approximate MOTS.  This surface ``just inside''
can be determined very easily by reducing the monopole term $F^{00}$ by
a small fraction.

\section{Summary}
\label{sec:summary} 
We have developed an apparent-horizon finder based on a multipole
expansion to arbitrary order $L$. The primary application we have in
mind is the numerical evolution of a binary black-hole system. In
order to check the performance of the MOTS finder in a spacetime of
similar structure we have performed careful tests using initial data
for two time-symmetric black holes.
 
From these tests it is evident that a reliable search for highly
distorted MOTS requires high order expansion. On the other hand,
using a high order expansion is very expensive and it is questionable
if this will be affordable during a dynamical evolution. 
 
However, in the evolution of a binary black-hole system for example,
it is desirable to detect a common MOTS as early as possible since
the region interior to this surface no longer needs to be evolved.
It may, therefore, be worthwhile searching for this common MOTS using
a high order expansion.
 
As a compromise, it is possible to use a low order
expansion for nearly spherical MOTS (as will be the case for the
MOTS around individual black holes or the common MOTS in the later
phases of an evolution) and a high order expansion for highly distorted
MOTS (as in the early phase of the common MOTS). Unfortunately, ``nearly
spherical'' is a coordinate dependent concept in this context, and
it is not clear how well the coordinates will behave in a
binary black-hole evolution code.
 
An obvious optimization of the code is to change to a minimization
scheme that uses derivatives. This is a fairly tedious, though
straightforward task.

\acknowledgments
We would like to thank Peter Anninos and Edward Seidel for helpful
discussions. This work was supported by NSF Grants AST 91-19475 and
PHY 94-08378, NASA Grant NAG-2809, and by the Grand Challenge grant
NSF PHY 93-18152 / ASC 93-18152 (ARPA supplemented).  Computations
were performed at the Cornell Center for Theory and Simulation in
Science and Engineering, which is supported in part by the National
Science Foundation, IBM Corporation, New York State, and the Cornell
Research Institute.

\appendix
\section{Useful expressions for STF tensors}
\label{app:expressions} 
The purpose of this appendix is to provide several useful expressions for 
spherical harmonics and STF tensors that have been omitted in the main
text.  All these formulae are taken from reference~\cite{thorne80}.
 
The spherical harmonics can be written
\begin{eqnarray}
Y^{\ell m}(\theta,\phi) &= C^{\ell m} e^{im\phi} P^{\ell m}(\cos\theta)
        \hspace*{1.7in} & \mbox{for $0 \leq m \leq \ell$} \nonumber\\
        &= C^{\ell m} (e^{i\phi} \sin\theta)^m
        {\displaystyle\sum_{j=0}^{[(\ell-m)/2]}}
	 a^{\ell mj}(\cos\theta)^{\ell-m-2j}
        \hspace{0.25in} & \mbox{for $0 \leq m \leq \ell$} \\
        &= (-1)^m Y^{\ell|m|*}
	\hspace*{2.16in} & \mbox{for $-\ell \leq m < 0$.}
	\nonumber
\end{eqnarray}
Here $*$ denotes complex conjugation, $[(\ell-m)/2]$ means the largest
integer less than or equal to $(\ell-m)/2$, and
\begin{eqnarray}
C^{\ell m} &=& (-1)^m \left( \frac{2\ell+1}{4\pi}
        \frac{(\ell-m)!}{(\ell+m)!} \right)^{1/2} \\
a^{\ell mj} &=& \frac{(-1)^j}{2^\ell j!(\ell-j)!}
	\frac{(2\ell-2j)!}{(\ell-m-2j)!}.
\end{eqnarray}
Note that the Cartesian components of a radial unit vector $n$ can
be written
\begin{equation}
n_x + in_y = e^{i\phi}\sin\theta,\;\;\; n_z = \cos\theta.
\end{equation}
In terms of these, the spherical harmonics are
\begin{equation}
Y^{\ell m}(\theta,\phi) = {\cal Y}^{\ell m}_{K_\ell} N_{K_\ell}(\theta,\phi),
\end{equation}
where the ${\cal Y}^{\ell m}_{K_\ell}$ are defined as
\begin{eqnarray} \label{basisSTF}
{\cal Y}^{\ell m}_{K_\ell} &= C^{\ell m} 
	{\displaystyle\sum_{j=0}^{[(\ell-m)/2]}} a^{\ell mj}
        (\delta^{1}_{(k_1} + i\delta^{2}_{(k_1}) \cdots
        (\delta^{1}_{(k_m} + i\delta^{2}_{(k_m}) \hspace{1.11in} & \nonumber\\ 
        & \hspace{0.95in}\mbox{}\times \delta^{3}_{k_{m+1}}
	\!\!\cdots\delta^{3}_{k_{\ell-2j}}
        (\delta^{a_1}_{k_{\ell-2j+1}} \delta^{a_1}_{k_{\ell-2j+2}})
        \cdots (\delta^{a_j}_{k_{\ell-1}} \delta^{a_j}_{k_{\ell})}) 
        \hspace{0.125in} & \mbox{for $0 \leq m \leq \ell$} \\
        &= (-1)^m ({\cal Y}^{\ell|m|}_{K_\ell})^* 
	\hspace{3.09in} & \mbox{for $-\ell \leq m < 0$}.\nonumber
\end{eqnarray}
$\delta^{a_i}_{k_\ell}$ is the Kronecker delta and the brackets
$A_{(k_1\cdots k_\ell)}$ denote complete symmetrization. Obviously, the
${\cal Y}^{\ell m}_{K_\ell}$ are completely symmetric by definition,
and they are also completely trace-free. Moreover, for every $\ell$ 
the $2\ell+1$ different ${\cal Y}^{\ell m}_{K_\ell}$ are linearly
independent and span a basis for the STF tensors of rank $\ell$. Any
STF tensor of rank $\ell$ can therefore be written as
\begin{equation}
{\cal F}_{K_\ell}=\sum_{m=-\ell}^{\ell} F^{\ell m}{\cal Y}^{\ell m}_{K_\ell}.
\end{equation}
For ${\cal F}_{K_\ell}$ to be real the coefficients have to satisfy
$F^{\ell-m} = (-1)^m (F^{\ell m})^*$.

\section{Storage}
\label{app:storage} 
Since we want to allow for expansion to arbitrary order $L$, we have 
to store objects of the type $N_{K_\ell}$ with multi-indices up to
arbitrary length $L$. This can be accomplished by noting that all these
objects are completely symmetric, i.e.
\begin{equation}
N_{k_1 \cdots k_i \cdots k_j \cdots k_\ell} = 
N_{k_1 \cdots k_j \cdots k_i \cdots k_\ell} 
\end{equation} 
for any pair of indices $k_i$ and $k_j$. In three dimensions these 
indices can only take the value $x$, $y$ or $z$. It is therefore
sufficient to specify the number of indices with value $x$, $y$ and 
$z$ to determine any element in $N_{K_\ell}$ uniquely. In the following
we adopt the notation
\begin{equation} \label{notation}
N_{(N_x)(N_y)(N_z)},
\end{equation}
where the $N_x$, $N_y$ and $N_z$ are the total numbers of indices $x$, 
$y$ and $z$. Obviously, the rank of the tensor is
\begin{equation}
\ell = N_x + N_y + N_z.
\end{equation}
A symmetric tensor of rank $\ell$ has
\begin{equation}
\sum_{i = 0}^\ell (i + 1) = \frac{1}{2} (\ell + 1)(\ell + 2)
\end{equation}
independent elements, the number of independent components for all
symmetric tensors of rank up to $L$ is
\begin{equation}
\sum_{\ell=0}^L \frac{1}{2} (\ell + 1)(\ell + 2) = 
\frac{1}{6} (L + 1)(L + 2)(L + 3).
\end{equation}
We can therefore store all symmetric tensors up to order $L$ in an 
array of this length. Each element is uniquely determined by a
combination of $N_x$, $N_y$ and $N_z$.
 
Note that an element $N_{(N_x)(N_y)(N_z)}$ appears
\begin{equation} \label{weighting}
\frac{\ell!}{N_x!N_y!N_z!}
\end{equation}
times in a symmetric tensor. This weighting factor has to be taken into 
account when carrying out sums as in~(\ref{stf}).
 
In our code we store $N_{K_\ell}$, its first and second derivative 
and the STF tensor ${\cal F}_{K_\ell}$ in arrays of this kind. In addition
to being symmetric the latter is completely trace-free, so that we 
could use even less storage. We decided not to do so, since this would
complicate the code for little benefit.
 
The next objects that need to be stored are the basis STF tensors
${\cal Y}^{\ell m}_{K_\ell}$. For each $\ell$ this requires storage of
$2\ell+1$ values.  We therefore need an array of length
\begin{equation} 
\sum_{\ell=0}^L \frac{1}{2} (2\ell + 1)(\ell + 1)(\ell + 2) = 
\frac{1}{12} (3L + 2)(L + 1)(L + 2)(L + 3).
\end{equation}
Each element is uniquely labeled by a combination of $N_x$, $N_y$ and 
$N_z$ together with an index $m$.
 
As a further complication these tensors are complex. However, since
${\cal Y}^{\ell(-m)}_{K_\ell} = (-1)^m ({\cal Y}^{\ell m}_{K_\ell})^*$ (see appendix A, 
equation~(\ref{basisSTF})), we need to store only the non-negative $m$.
This can be accomplished by storing the real parts in the storage for
$m \ge 0$ and the imaginary parts in the storage for $m < 0$. 
 
A sum as in~(\ref{sum}) can be written 
\begin{eqnarray}
\label{STFrealsum}
{\cal F}_{K_\ell} &=& \sum_{m=-\ell}^{\ell} F^{\ell m}
	{\cal Y}^{\ell m}_{K_\ell} \nonumber\\
        &=& F^{\ell0} {\cal Y}^{\ell0}_{K_\ell} + \sum_{m = 1}^{\ell} \left[ 
        F^{\ell m} {\cal Y}^{\ell m}_{K_\ell} + 
	(F^{\ell m} {\cal Y}^{\ell m}_{K_\ell})^* \right] \\
        &=& F^{\ell0} {\cal Y}^{\ell0}_{K_\ell} + 2 \sum_{m = 1}^{\ell}
        \left[ \Re F^{\ell m} \Re {\cal Y}^{\ell m}_{K_\ell}  -
        \Im F^{\ell m} \Im {\cal Y}^{\ell m}_{K_\ell} \right] \nonumber
\end{eqnarray}
(where we have assumed $F^{\ell-m} = (-1)^m (F^{\ell m})^*$ so that 
${\cal F}_{K_\ell}$ is real, see appendix A). Storing the elements as described
above, we can again sum from $-\ell$ to $\ell$, but we have to take into
account a weight of $2$ for $m>0$ and $-2$ for  $m<0$, as given by
eqn.~(\ref{STFrealsum}).
 
The last objects that have to be stored are the expansion coefficients
$F^{\ell m}$. Since for each $\ell$ there are $2\ell + 1$ different elements,
we have to store a total of
\begin{equation} \label{flm}
\sum_{\ell=0}^{L} (2\ell + 1) = (L + 1)^2
\end{equation}
elements. Again, these elements will be complex and satisfy
$F^{\ell-m} = (-1)^m (F^{\ell m})^*$, so that we can again store the real 
parts in the storage for $m \ge 0$ and the imaginary parts in the
storage for $m < 0$.

\section{Initialization and recurrence relations}
\label{app:init_recur}
Before a MOTS can be found $N_{K_\ell}$, its derivatives and
${\cal Y}^{\ell m}_{K_\ell}$ need to be initialized for various directions
$(\theta,\phi)$. Although direct expressions could be used to do so
(as for example equation~(\ref{basisSTF}) in Appendix A), this would
be extremely tedious and inefficient.  Moreover, sums like
(\ref{basisSTF}) are prone to cancellation error from adding terms of
opposite sign.  We have therefore derived recursive expressions that
make the initialization much easier.
 
Since the ${\cal Y}^{\ell m}_{K_\ell}$ are closely related to the spherical
harmonics, we can start with the standard recurrence relation
\begin{equation}
Y^{\ell m} = \sqrt{ \frac{2\ell+1}{\ell^2 - m^2} }
        \Biggl( \sqrt{2\ell-1}\, Y^{(\ell-1)m}\cos \theta - 
        \sqrt{ \frac{(\ell-1)^2 - m^2}{2\ell-3}}\, Y^{(\ell-2)m} \Biggr).
\end{equation}
Here we can use $\cos \theta = n_z$, insert 
$n_x^2 + n_y^2 + n_z^2 = 1$, and replace the spherical harmonics 
with equation~(\ref{ylm}), which yields
\begin{equation}
\label{STFrec1}
{\cal Y}^{\ell m}_{K_\ell} N_{K_\ell} =
        \sqrt{ \frac{2\ell+1}{\ell^2 - m^2} } \Biggl( \sqrt{2\ell-1}\,
	{\cal Y}^{(\ell-1)m}_{K_{\ell-1}} N_{K_{\ell-1}} n_z
        - \sqrt{ \frac{(\ell-1)^2 - m^2}{2\ell-3}}\,
        {\cal Y}^{(\ell-2)m}_{K_{\ell-2}} N_{K_{\ell-2}}
	(n_x^2 + n_y^2 + n_z^2) \Biggr).
\end{equation}
Adopting our notation (\ref{notation}), we can rewrite this equation as
\begin{eqnarray}
\lefteqn{\frac{\ell!}{N_x!N_y!N_z!}
	{\cal Y}^{\ell m}_{(N_x)(N_y)(N_z)} N_{(N_x)(N_y)(N_z)} =
        \sqrt{ \frac{2\ell+1}{\ell^2 - m^2} }
        \Biggl[} \\
	& & \hspace{0.4in} \frac{(\ell-1)!}{N_x!N_y!(N_z-1)!} 
        \sqrt{2\ell-1}\, {\cal Y}^{(\ell-1)m}_{(N_x)(N_y)(N_z-1)} -
        \sqrt{ \frac{(\ell-1)^2 - m^2}{2\ell-3}} \biggl(
         \frac{(\ell-2)!}{(N_x-2)!N_y!N_z!} 
        {\cal Y}^{(\ell-2)m}_{(N_x-2)(N_y)(N_z)} \nonumber\\
        & & \hspace{0.45in} \mbox{} + \frac{(\ell-2)!}{N_x!(N_y-2)!N_z!} 
        {\cal Y}^{(\ell-2)m}_{(N_x)(N_y-2)(N_z)} + 
        \frac{(\ell-2)!}{N_x!N_y!(N_z-2)!}       
        {\cal Y}^{(\ell-2)m}_{(N_x)(N_y)(N_z-2)} \biggr) 
        \Biggr] N_{(N_x)(N_y)(N_z)}. \nonumber
\end{eqnarray}
Note that, because the ${\cal Y}^{\ell m}_{N_{K_\ell}}$ are
direction-independent, only specific terms in the STF tensor sums are
allowed on the right-hand side of eq.~(\ref{STFrec1}).  Only those
terms that, together with the additional unit vectors appearing on the
right-hand size of eq.~(\ref{STFrec1}), result in the same number of
$N_x$, $N_y$, and $N_z$ that appear on its left-hand size are allowed.
Finally, the direction-independence of the
${\cal Y}^{\ell m}_{N_{K_\ell}}$ means that we can drop the the
$N_{(N_x)(N_y)(N_z)}$ terms (which contain all of the directional
dependence) to get
\begin{eqnarray}
{\cal Y}^{\ell m}_{(N_x)(N_y)(N_z)} &=& 
	\sqrt{ \frac{2\ell+1}{\ell^2 - m^2} }
        \Biggl[ \frac{N_z}{\ell} \sqrt{2\ell-1}
        {\cal Y}^{(\ell-1)m}_{(N_x)(N_y)(N_z-1)} \nonumber\\
	&& \hspace{0.5in} \mbox{}- \frac{1}{\ell(\ell-1)}
	\sqrt{ \frac{(\ell-1)^2 - m^2}{2\ell-3}} \biggl(
	N_x(N_x-1) {\cal Y}^{(\ell-2)m}_{(N_x-2)(N_y)(N_z)} \\
	&& \hspace{1.0in} \mbox{}
	+ N_y(N_y-1) {\cal Y}^{(\ell-2)m}_{(N_x)(N_y-2)(N_z)} +
        N_z(N_z-1) {\cal Y}^{(\ell-2)m}_{(N_x)(N_y)(N_z-2)} \biggr) 
        \Biggr]. \nonumber
\end{eqnarray}
This expression can be used to initialize the ${\cal Y}^{\ell m}_{K_\ell}$ with
$-\ell < m < \ell$. For $m = \ell$ we can use eq.~(\ref{basisSTF}), which 
for this case reduces to
\begin{equation}
{\cal Y}^{\ell\ell}_{(N_x)(N_y)(N_z)} = \left\{ 
        \begin{array}{ll}
        0 & \mbox{if $N_z \neq 0$} \\
        \displaystyle \frac{(-1)^\ell i^{(N_y)}}{2^\ell \ell!} 
        \sqrt{ \frac{(2\ell +1)(2\ell)!}{4 \pi}} & \mbox{if $N_z = 0$}
        \end{array} \right.
\end{equation}
 
Next we have to initialize the tensors $N_{K_\ell}$ (eq.~(\ref{nkl})) as
well as their first and second derivatives. Note that the partial 
derivative of a unit vector is
\begin{equation}
\partial_i n_j = \frac{1}{r} \left(\delta_{ij} - n_i n_j \right)
\equiv n_{ij}.
\end{equation}
The partial derivative of $N_{K_\ell}$ is therefore
\begin{equation} \label{firstd}
\partial_i N_{K_\ell} =
\partial_i (n_{k_1} n_{k_2} \cdots n_{k_\ell} ) =
\ell \,n_{i(k_\ell} N_{K_{\ell-1})}.
\end{equation}
Taking a second derivative yields, after some algebra,
\begin{equation} \label{secondd}
\partial_i \partial_j N_{K_\ell} = - \frac{\ell}{r} \bigl(n_{ij} N_{K_\ell}
	+ n_i n_{j(k_\ell} N_{K_{\ell-1})}
	+ n_j n_{i(k_\ell} N_{K_{\ell-1})} \bigr)
	+ \ell(\ell-1) n_{i(k_{\ell-1}} N_{K_{\ell-2}} n_{k_\ell)j}.
\end{equation}
Note that we can now construct $N_{K_\ell}$, $\partial_i N_{K_\ell}$ and
$\partial_i \partial_j N_{K_\ell}$ from the ten different,
totally-symmetric objects $N_{K_\ell}$, $n_{i(k_\ell} N_{K_{\ell-1})}$ and 
$n_{i(k_{\ell-1}} N_{K_{\ell-2}} n_{k_\ell)j}$. Denoting any one of these
with $S_{(N_x)(N_y)(N_z)}$ (in our notation~(\ref{notation})) we find
that they all satisfy the recurrence relation
\begin{equation}
S_{(N_x)(N_y)(N_z)} =
	\frac{ N_x n_x S_{(N_x-1)(N_y)(N_z)}
	+ N_y n_y S_{(N_x)(N_y-1)(N_z)}
        + N_z n_z S_{(N_x)(N_y)(N_z-1)} } {N_x + N_y + N_z} .
\end{equation}
The starting values for the different objects are
\begin{equation}
\begin{array}{cccc}
&N_{K_\ell}&n_{a(k_\ell}N_{K_{\ell-1})} 
        & n_{a(k_{\ell-1}}N_{K_{\ell-2}}n_{k_{\ell-1})b}  \\[3mm]
l=0     & 1     & 0                    
        & 0                                     \\[3mm] 
l=1     & n_i   & n_{ai}               
        & 0                                     \\[3mm]
l=2     & n_in_j& \frac{1}{2}(n_{ai}n_j + n_{aj}n_i) 
        & \frac{1}{2} (n_{ai}n_{bj} + n_{aj}n_{bi}).
\end{array}
\end{equation}
Once these objects have been calculated up to order $L$, we can then 
construct the derivatives of $N_{K_\ell}$ using eqs.~(\ref{firstd})
and~(\ref{secondd}).

\section{The area Element on the apparent horizon}
\label{app:area_elem}
Beginning with the line element for the three-dimensional spatial
hypersurface
\begin{equation}\label{eqn:3_line_el}
ds^2 = g_{ij}dx^idx^j,
\end{equation}
we transform to polar coordinates centered around some point using
\begin{equation}\label{eqn:polarcoords}
dx^i = \frac{\partial x^i}{\partial r} dr +
        \frac{\partial x^i}{\partial \theta} d\theta +
        \frac{\partial x^i}{\partial \phi} d\phi.
\end{equation}
The line element for the surface parameterized by $r = f(\theta,\phi)$
is obtained by substituting
\begin{equation}
dr = \frac{\partial f}{\partial \theta} d\theta +
        \frac{\partial f}{\partial \phi} d\phi
\end{equation}
into eqn.~(\ref{eqn:polarcoords}).  We define
\begin{equation}
\Theta^i = \frac{1}{r} \left( \frac{\partial x^i}{\partial r}
        \frac{\partial f}{\partial \theta} + 
        \frac{\partial x^i}{\partial \theta} \right)
\end{equation}
and 
\begin{equation}
\Phi^i =  \frac{1}{r \sin\theta} \left( \frac{\partial x^i}{\partial r}
        \frac{\partial f}{\partial \phi} +
        \frac{\partial x^i}{\partial \phi} \right),
\end{equation}
which can be expanded to yield
\begin{eqnarray}
\Theta^x &=& \frac{n_xn_z}{\sqrt{1-n_z^2}}(1 + n_x\partial_xf
	+ n_y\partial_yf) - n_x\sqrt{1-n_z^2}\partial_zf, \\
\Theta^y &=& \frac{n_yn_z}{\sqrt{1-n_z^2}}(1 + n_x\partial_xf
	+ n_y\partial_yf) - n_y\sqrt{1-n_z^2}\partial_zf, \\
\Theta^z &=& \frac{n_z^2}{\sqrt{1-n_z^2}}(n_x\partial_xf
	+ n_y\partial_yf) - \sqrt{1-n_z^2}(1 + n_z\partial_zf), \\
\Phi^x &=& \frac{1}{\sqrt{1-n_z^2}}[n_x(n_x\partial_yf
	- n_y\partial_xf) - n_y], \\
\Phi^y &=& \frac{1}{\sqrt{1-n_z^2}}[n_y(n_x\partial_yf
	- n_y\partial_xf) + n_x], \\
\Phi^z &=& \frac{1}{\sqrt{1-n_z^2}}(n_x\partial_yf - n_y\partial_xf).
\end{eqnarray}
Finally, we can now rewrite the line element (\ref{eqn:3_line_el}) as
\begin{equation}
ds^2 = f^2 \left( \Theta^i \Theta_i d\theta^2
        + 2 \Theta^i \Phi_i \sin\theta d\theta d\phi
        + \Phi^i \Phi_i \sin^2 \theta d\phi^2 \right),
\end{equation}
where we have lowered the indices on $\Theta$ and $\Phi$ with the
metric $g_{ij}$. From this we find the two-surface area element
\begin{equation}
\label{area_elem}
d^2\sigma = \sqrt{ (\Theta^i \Theta_i)
        (\Phi^j \Phi_j ) - (\Theta^k \Phi_k)^2 } f^2 \sin \theta
        d\theta d\phi.
\end{equation}

%
%
%  REFERENCES
%
%

\end{document}